\documentclass{appolb}
\usepackage{graphicx}
\usepackage{color}
\usepackage{amsmath}

\begin{document}
\title{\textit{Ab initio} optical potentials and nucleon scattering on medium mass nuclei%
\thanks{Presented at the Zakopane Conference on Nuclear Physics ``Extremes of the Nuclear
Landscape'', Zakopane, Poland, August 28 -- September 4, 2016.}%
}
\author{A. Idini$^1$,  C. Barbieri$^1$,  P. Navr\'atil$^2$
\address{$^1$Department of Physics, University of Surrey, Guildford, GU2 7XH, United Kingdom\\
         $^2$TRIUMF, 4004 Wesbrook Mall, Vancouver, British Columbia, V6T 2A3, Canada}
}
         
\maketitle

\begin{abstract}
We show the first results for the   elastic scattering of neutrons off oxygen and calcium isotopes obtained from  {\em ab initio} optical potentials. The  potential is derived using self consistent Green's function theory (SCGF) with the saturating chiral interaction NNLO$_{\textrm{sat}}$.
Our calculations are compared to available scattering data and show that it is possible to reproduce low energy scattering observables in medium mass nuclei  from first principles.
\end{abstract}
  
\section{Introduction}
Recent years have seen considerable advances in the theory of optical potentials. Non locality effects have been seen to be necessary for describing three-body processes~\cite{Timofeyuk:13,Bailey:16}, the importance of both scattering and bound states in the coupling to breakup channels has been  explored~\cite{Garcia:15}, and global dispersive optical potentials have been developed~\cite{Dickhoff:16}. 

The greatest challenge remains, however, the one of describing the nuclear structure and scattering consistently, from the same theory.
Many-body Green's function methods are particularly suited to attempt this for medium and large nuclei since their central quantity, the self-energy, is naturally linked to the Feshbach theory of optical potentials~\cite{Feshbach:58,Jeukenne:77}. In particular, the particle part of the self-energy is equivalent to the original formulation of Feshbach, while its hole part describes the structure of the target~\cite{Escher:02}.
Nuclear field theory is one of the first  (semi  phenomenological)  attempts to build such a theory for atomic nuclei \cite{Mahaux:85,Idini:12} and it has been extended to nuclear transfer reactions \cite{Idini:15,Broglia:16}. Another incarnation of Green's function related theories is the dispersive optical model \cite{Johnson:88}, which is a data driven formulation of global  (local and non local) potentials constructed as the best possible parameterization of the microscopic self-energy~\cite{Charity:06,Dickhoff:16}.

For transfer reactions, such as $(d,p)$,  it would be particularly important to have an optical potential that is deduced consistently from the same Hamiltonian used in the proton-neutron channel~\cite{Bailey:16}. To do so, one  needs  {\em realistic} nuclear interactions and {\em ab initio} calculations of elastic nucleon-nucleus scattering.
The no-core shell model with continuum (NCSMC) has been successful to calculate scattering and transfer reactions for light targets~\cite{Navratil:10, Baroni:13prl, Raimondi:16}. On the other hand, the  self consistent Green's function (SCGF) formalism~\cite{Dickhoff:04,Soma:11} is better suited to derive optical potentials for medium mass targets. SCGFs have been used to calculate phase shifts~\cite{Barbieri:05} and to investigate analytical properties  of optical potentials~\cite{Waldecker:11}. However,  these calculations were limited to two-body forces and a direct comparison to the experiment has been hindered by the lack of realistic interactions capable to reproduce accurately nuclear radii.

Three-body interactions have been recently implemented for SCGF in~\cite{Carbone:13,Cipollone:13,Cipollone:15}. Moreover, the introduction of the NNLO$_{\textrm{sat}}$ interaction~\cite{Ekstrom:15} has allowed a good reproduction of nuclear saturation and, hence, of radii and binding energies across the oxygen~\cite{Lapoux:16} and calcium chains~\cite{Garcia:16}. Although this interaction has limitations regarding the symmetry energy in neutron rich nuclei, we are now in the position to make a meaningful comparison of first principle approaches  to scattering data. Here, we perform state of the art SCGF calculations to test the quality of current {\em ab initio}  methods and  of the NNLO$_{\textrm{sat}}$ Hamiltonian in predicting elastic scattering.

\section{The microscopic optical potential}
The irreducible self-energy, $\Sigma^\star(\omega)$, has the general spectral representation
\begin{align}
\Sigma^\star_{\alpha \beta}(E) = \Sigma^{(\infty)}_{\alpha \beta} ~+~  & \sum_{i, j}  {\bf M}^{\dagger}_{\alpha, i} \left[ \frac{1}{E - (\textbf{K}^> + \textbf{C}) + i \eta} \right]_{i, j} {\bf M}_{j,\beta}
 \nonumber \\ 
 & \quad  + ~\sum_{r, s} {\bf N}_{\alpha, r} \left[ \frac{1}{E - (\textbf{K}^< + \textbf{D}) - i \eta} \right]_{r, s} {\bf N}^{\dagger}_{s,\beta} \; ,
\end{align}
where $\Sigma^{(\infty)}_{\alpha \beta}$ is the correlated and energy independent mean field. 
We perform calculations with the third order algebraic diagrammatic construction [ADC(3)] method, where
 the matrices  $\bf{M}$~($\bf{N}$)  couple single particle states to intermediate 2p1h~(2h1p) configurations, ${\bf C}$~(${\bf D}$) are interaction matrices among these configurations and $\bf{K}$ are their unperturbed energies.

We use a spherical harmonic oscillator basis consisting of $N_{\rm max}$+$1$ oscillator shells, so the optical potential for a given
partial wave  ($l,j$) is expressed in terms of the oscillator radial functions $R_{n,l}(r)$ as
\begin{equation}
\Sigma^{\star  \,  l,j}(r, r'; E) = \sum_{n,n'} R_{n,l}(r) \,  \Sigma^{\star \, l,j}_{n,n'}(E)  \, R_{n',l}(r') \; ,
\label{eq:sigma}
\end{equation}
which is non local and depends on energy, angular momentum and parity.
We solve the corresponding scattering problem in the full one-body space (so that  the kinetic energy is treated exactly, without truncations) and account for the non locality and $l,j$ dependence of Eq.~\eqref{eq:sigma}. For each partial wave and parity, the phase shifts $\delta(E)$ are obtained as function of the projectile energy, from where the differential cross section is calculated.
We show results for incident energies in the laboratory frame, except for Fig.~\ref{Fig:dSig_nCa40} below.

\begin{figure}[t]
\centerline{%
\includegraphics[width=0.45\textwidth]{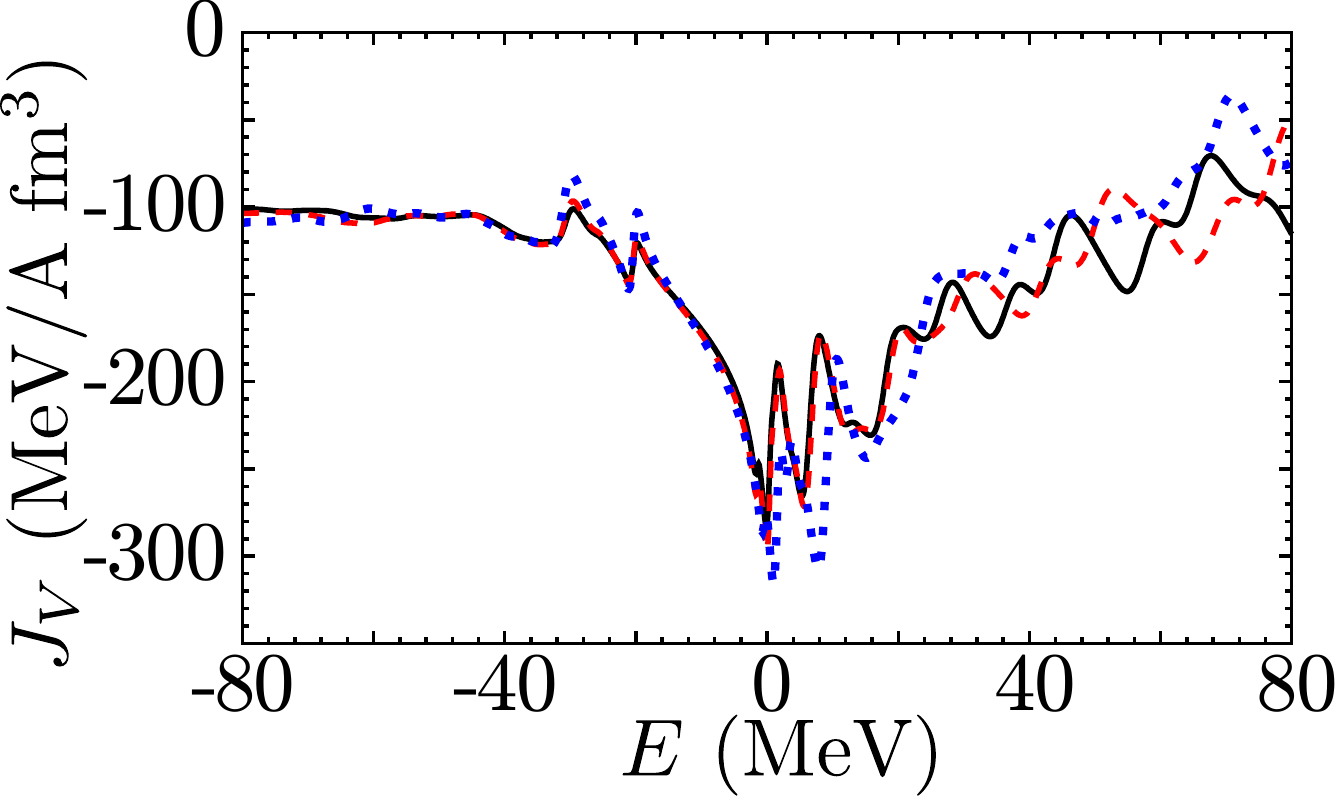} \includegraphics[width=0.45\textwidth]{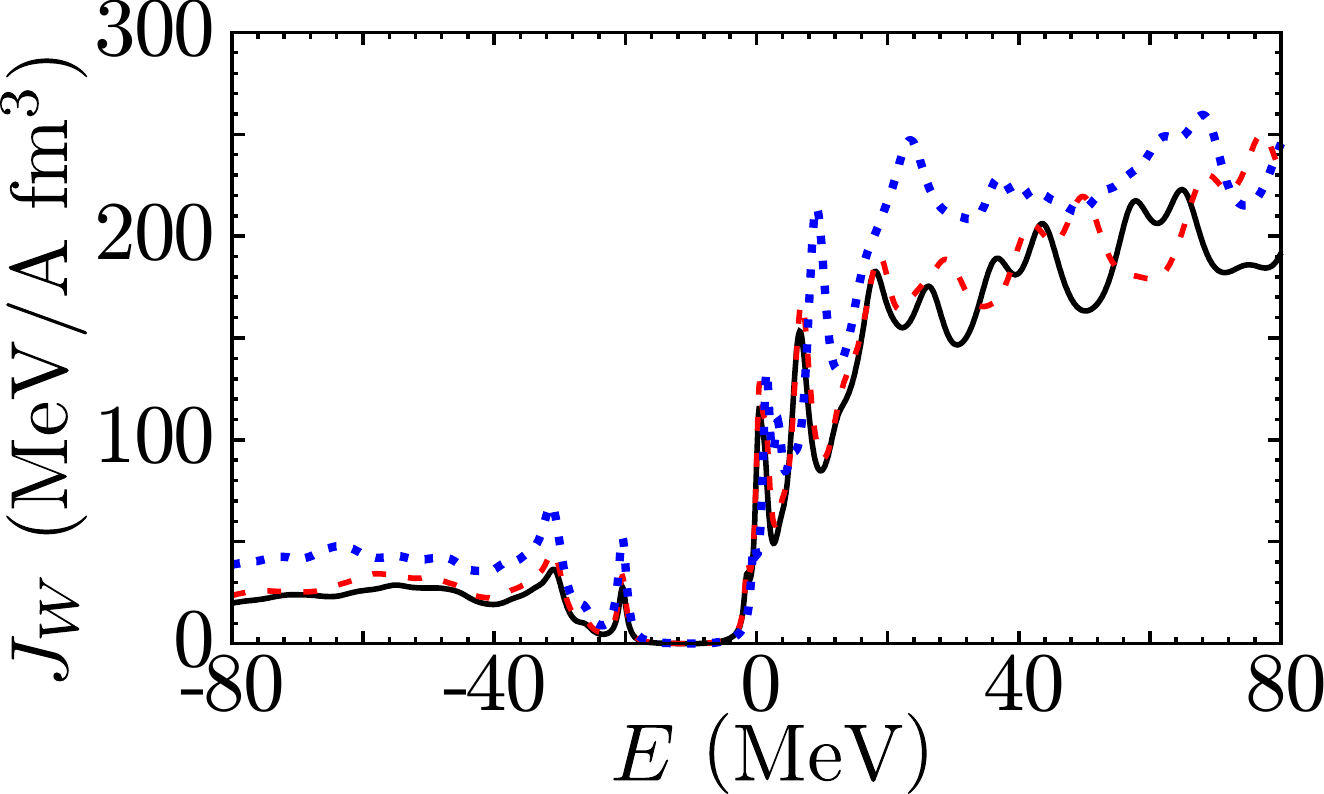}}
\caption{Volume integrals of the real ({\em left}) and imaginary ({\em right}) parts of the \hbox{neutron-$^{16}$O} optical potential calculated for different numbers oscillator shells in the model space: $N_{\rm max} = 7$ (dotted), 11 (dashed) and 13 (solid lines). Note that $\Im m\{\Sigma^\star(E$=$E_F)$=0, so $J_W (E$=$E_F)$=0, where $E_F$ is the Fermi energy. Thus, the potential for particle (holes) states is above (below) the gap in the $J_W$ plot.}
\label{Fig:1}
\end{figure}

\section{Results}
In the following, we consider the volume integrals of the real and imaginary parts of the self-energy (i.e., the optical potential):
\begin{eqnarray}
 J_V (E) = &  4 \pi \int \textrm{d}r r^2 \int \textrm{d}r' r'^2 \sum_{l,j} \Re e\{ \Sigma^{\star \, l,j} (r,r'; E) \} \; , \\
 J_W (E) = &  4 \pi \int \textrm{d}r r^2 \int \textrm{d}r' r'^2 \sum_{l,j}  \Im m\{ \Sigma^{\star \, l,j} (r,r'; E) \} \; ,
\end{eqnarray}
since these are  strongly constrained by  experimental data~\cite{Jeukenne:77}.

Fig. \ref{Fig:1}  shows the  volume integrals of the neutron-$^{16}$O for different model space truncations. 
Both the part of the self-energy below the Fermi surface (which describes the structure of the target) and the resonant structures
for scattering at low energy are substantially converged already for $N_{\rm max}$=11.
The oscillations seen at higher energies ($E>$10~MeV) are an artefact of using a discretized model space and  keep changing with $N_{\rm max}$. They can fade away for an infinite  space, or by exploiting a basis with the continuum.

\begin{figure}[t]
\centerline{%
\includegraphics[width=0.80\textwidth]{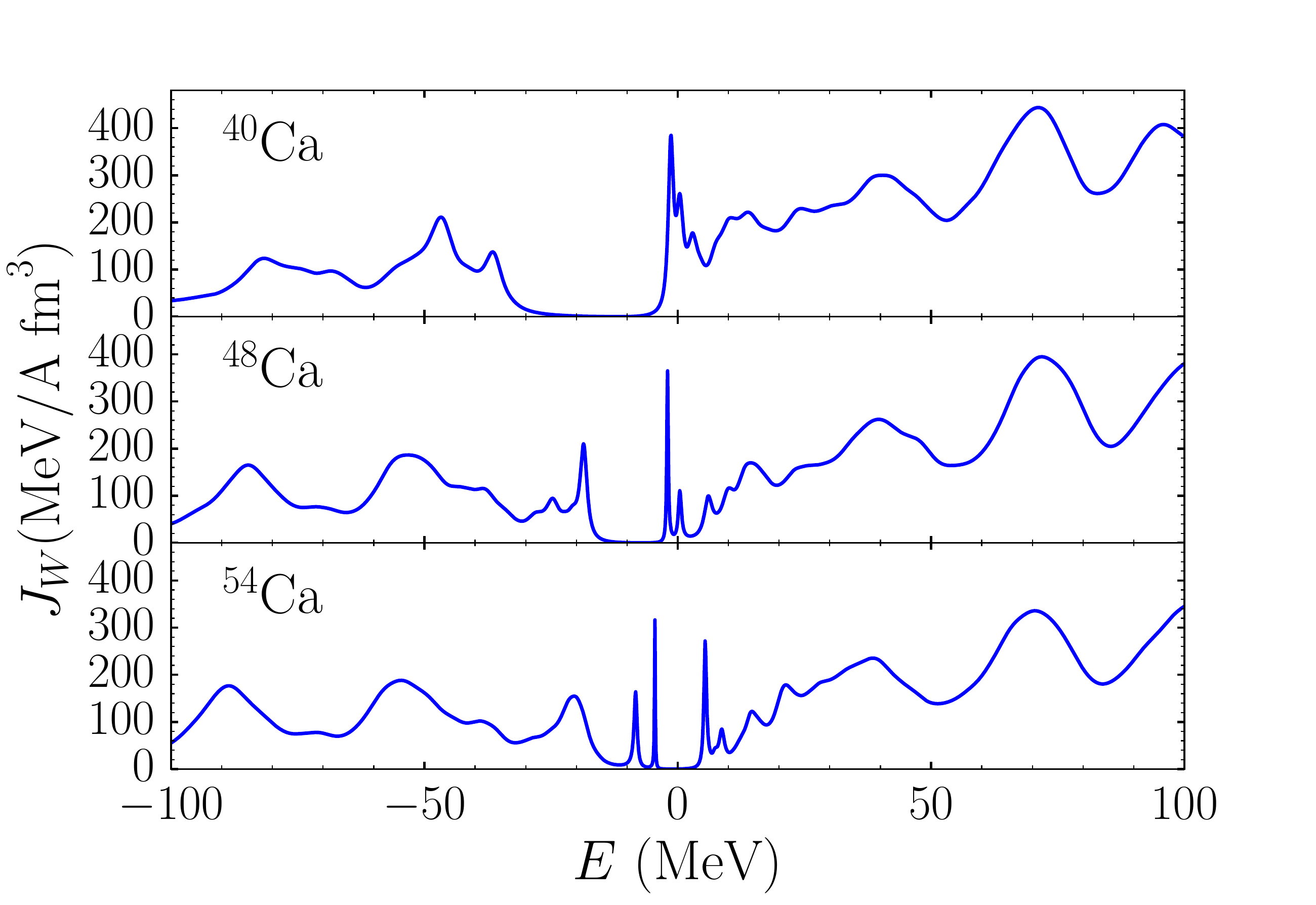} }
\caption{Volume integral of the imaginary part of the neutron optical potentials, $J_W (E)$, for the $^{40}$Ca, $^{48}$Ca and $^{54}$Ca targets. Calculated at $N_{\rm max}$=11.}
\label{Fig:Jw_Ca}
\end{figure}

Fig.~\ref{Fig:Jw_Ca} shows $J_W$ for selected closed sub-shell Ca isotopes. The gap at the Fermi
surface, where $\Im m\Sigma^\star(E)$=0, shifts to higher energies and eventually crosses the continuum threshold with increasing neutron number. Compared to previous calculations using the Argonne $v_{18}$ and N$^3$LO(500) interactions~\cite{Waldecker:11}, the NNLO$_{\textrm{sat}}$ predicts an increased level density  in the proximity of the Fermi energy, as expected for a correct nuclear saturation.

%

\begin{figure}[t]
\centerline{%
\includegraphics[width=0.50\textwidth]{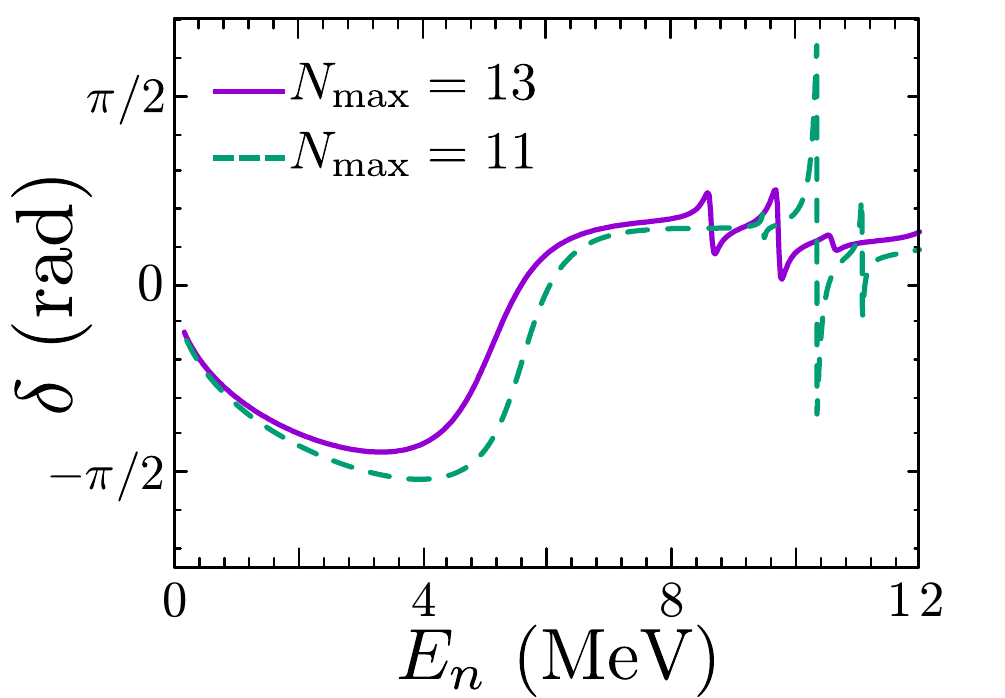} \hspace{0.02\textwidth}
\includegraphics[width=0.44\textwidth]{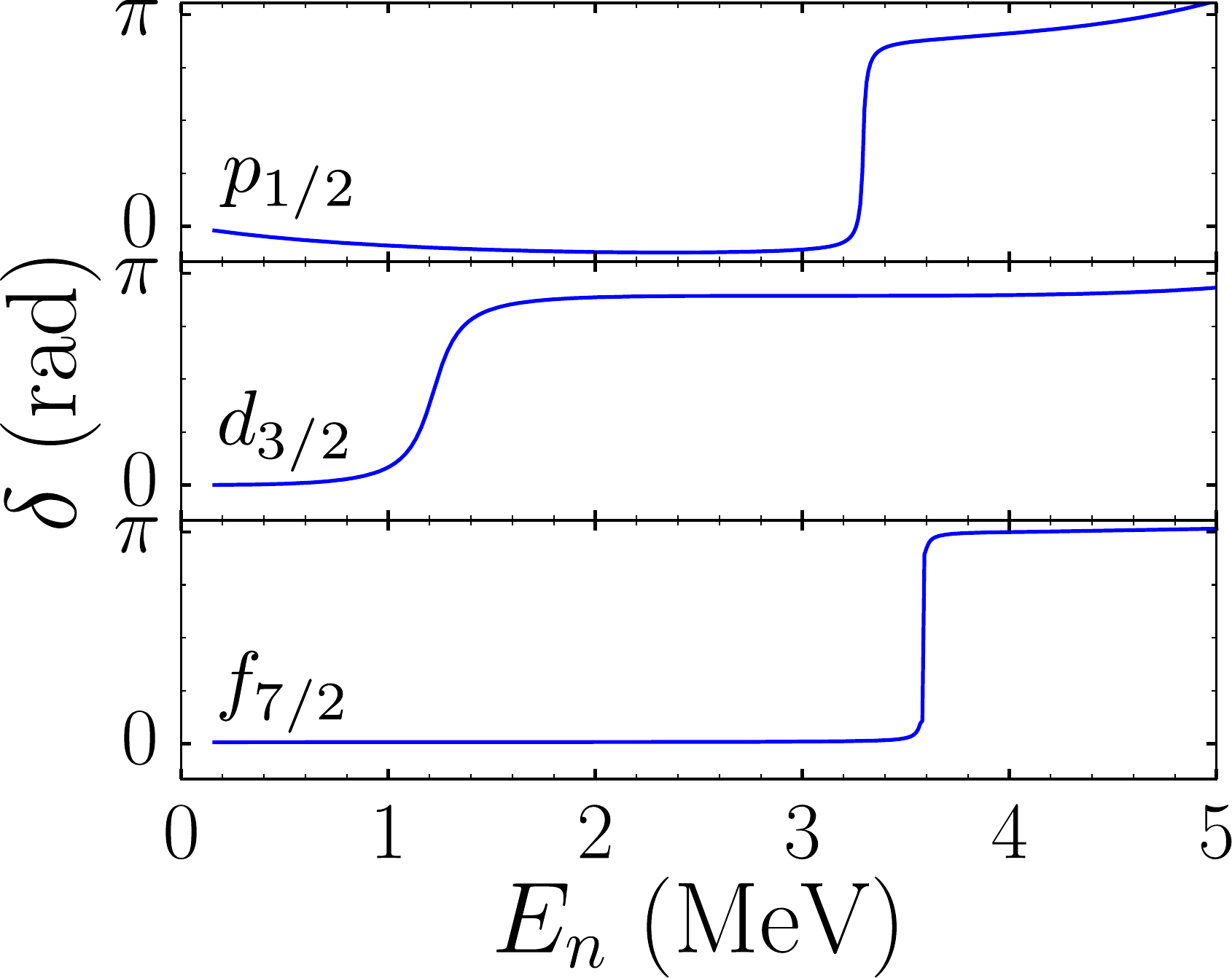}
}
\caption{Nuclear phase shifts, $\delta(E)$, for scattering off $^{16}$O as a function of the incident neutron energy.
{\em Left panel}: dependence of the  $s_{1/2}$ partial wave on the number of oscillator shells, for $N_{\rm max}$=11 (dashed) and 13 (solid). The oscillations at larger energies are narrow resonances. 
{\em Right panel}: $d_{3/2}$, $p_{1/2}$ and  $f_{7/2}$ partial waves.
}
\label{Fig:phsh}
\end{figure}

In Fig.~\ref{Fig:phsh}, the neutron $s_{1/2}$ phase shifts for  $^{16}$O  is shown for $N_{\rm max}$=11 and~13. 
The  resonance at $E \approx$5~MeV is almost converged for these spaces. Note that this state is dominated by 2p1h components and thus it can still be affected by many-body truncations.  The wiggles computed at energies $E>$8~MeV are due to similar but very narrow resonances. Again, these are linked  to the discretisation of the model space and drift when increasing the number of oscillator shells.
%
%
The right panel of Fig.~\ref{Fig:phsh} shows the phase shifts for other representative partial waves.
The $p_{1/2}$ has a sub-threshold bound state, while there is a very narrow $f_{7/2}$ that is also seen experimentally within 1~MeV of our calculation~\cite{Lister:66}. The dominant $d_{3/2}$ resonance is converged with respect to $N_{\rm max}$ and it is computed at $\approx$1.15~MeV in the c.o.m. energy, while the experimental value is 0.94~MeV.
 In general, we find that NNLO$_{sat}$  predicts the location of dominant quasiparticle and holes states with a (conservative) accuracy of $<$2~MeV for this nucleus.

\begin{figure}[t]
\centerline{%
\includegraphics[width=0.5\textwidth]{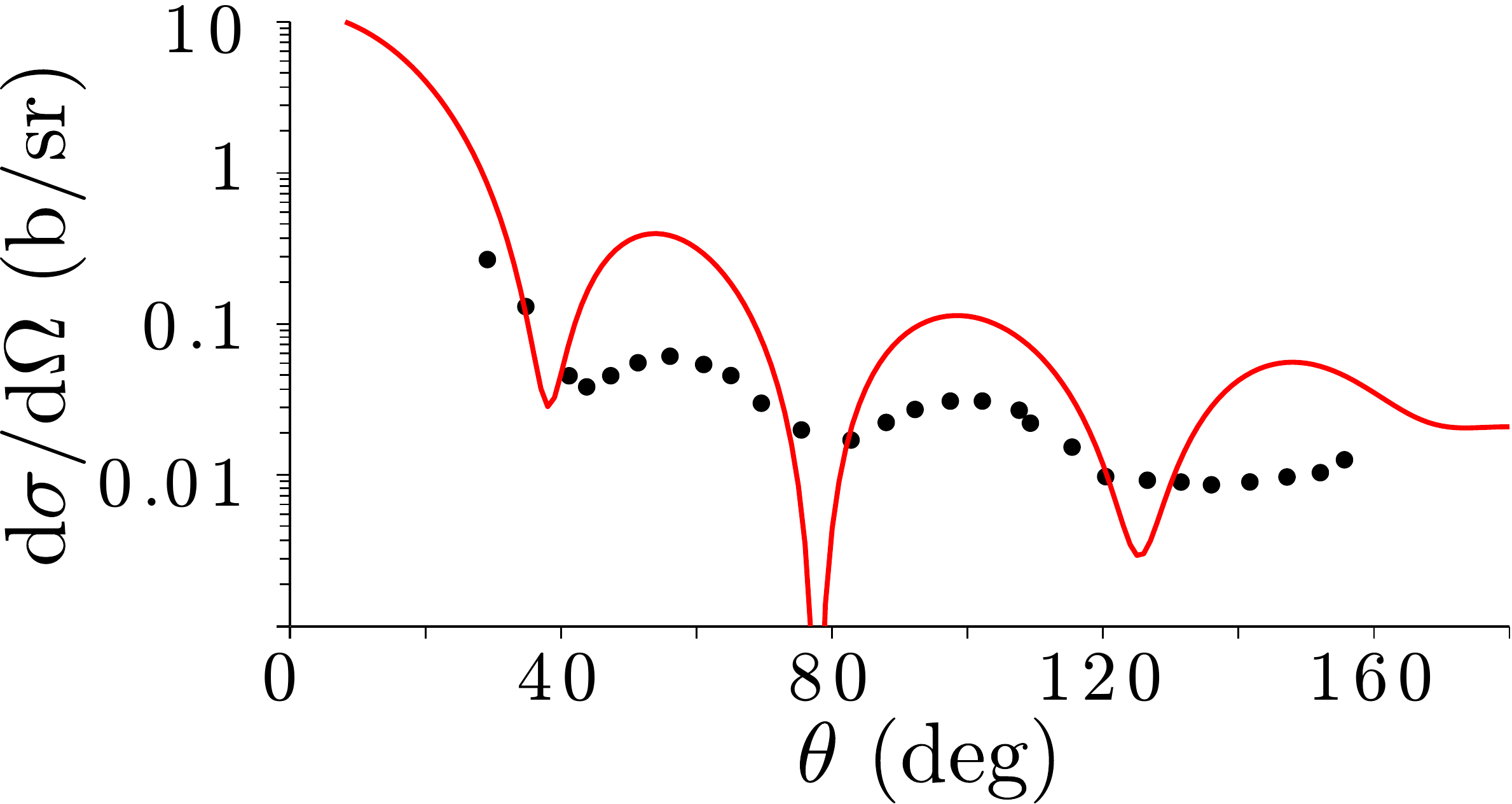}}
\caption{Plot of differential  cross section for neutron elastic scattering on $^{40}$Ca at 13.56 MeV of center of mass energy  compared with experimental data from~\cite{Honore:86}.  
Note that proton scattering on $^{40}$Ca was similarly computed in Ref.~\cite{Hagen:12}
}
\label{Fig:dSig_nCa40}
\end{figure}

Finally, Fig.~\ref{Fig:dSig_nCa40} compares the differential cross section for the elastic scattering of neutrons off $^{40}$Ca with  the experiment at 13.56 MeV c.o.m. energy, with $N_{\rm max}$=11.  Minima in the cross section are reproduced reasonably well, confirming the correct prediction of matter radii, but there appears to be a general lack of absorption.  This may be due to either missing doorway configurations (3p2h and beyond) or to the (still crude)  model space.

Even with the limitations of a (non optimal) oscillator basis, we found that most important features of optical potentials are well reproduced. In the long term, it will be necessary to properly account for the continuum in calculating the self-energy and to improve the realistic nuclear interactions. Nevertheless, it is  clear from the present results that reliable {\em ab initio} calculations of optical potentials are now a goal within reach.





\bibliographystyle{polonica}

\bibliography{nuclear.bib}

\begin{thebibliography}{10}

\bibitem{Timofeyuk:13}
N.~K. Timofeyuk, R.~C. Johnson,
\newblock {\em Phys. Rev. Lett.}
\newblock {\bf  110}, 112501 (2013).

\bibitem{Bailey:16}
G.~W. Bailey, N.~K. Timofeyuk, J.~A. Tostevin,
\newblock {\em ibid.}
\newblock {\bf  117}, 162502 (2016).

\bibitem{Garcia:15}
Fern\'andez-Garc\'{\i}a {\em et al.},
\newblock {\em Phys. Rev. C}
\newblock {\bf  92}, 054602 (2015).

\bibitem{Dickhoff:16}
W.~H. Dickhoff, R.~J. Charity, M.~H. Mahzoon,
\newblock
\newblock {\em arXiv:1606.08822 [nucl-th]} (2016).

\bibitem{Feshbach:58}
H.~Feshbach,
\newblock {\em Annals of Physics}
\newblock {\bf  5}, 357  (1958).

\bibitem{Jeukenne:77}
J.-P. Jeukenne, A.~Lejeune, C.~Mahaux,
\newblock {\em Phys. Rev. C}
\newblock {\bf  16}, 80 (1977).

\bibitem{Escher:02}
J.~Escher, B.~K. Jennings,
\newblock {\em ibid.}
\newblock {\bf  66}, 034313 (2002).

\bibitem{Mahaux:85}
C.~Mahaux, P.~F. Bortignon, R.~A. Broglia, C.~H. Dasso,
\newblock {\em Phys. Rep.}
\newblock {\bf  120}, 1 (1985).

\bibitem{Idini:12}
A.~Idini, F.~Barranco, E.~Vigezzi,
\newblock {\em Phys. Rev. C}
\newblock {\bf  85}, 014331 (2012).

\bibitem{Idini:15}
A.~Idini, G.~Potel, F.~Barranco, E.~Vigezzi, R.~A. Broglia,
\newblock {\em ibid.}
\newblock {\bf  92}, 031304 (2015).

\bibitem{Broglia:16}
R.~A. Broglia, P.~F. Bortignon, F.~Barranco, E.~Vigezzi, A.~Idini, G.~Potel,
\newblock {\em Physica Scripta}
\newblock {\bf  91}, 063012 (2016).

\bibitem{Johnson:88}
C.~H. Johnson, C.~Mahaux,
\newblock {\em Phys. Rev. C}
\newblock {\bf  38}, 2589 (1988).

\bibitem{Charity:06}
R.~J. Charity, L.~G. Sobotka, W.~H. Dickhoff,
\newblock {\em Phys. Rev. Lett.}
\newblock {\bf  97}, 162503 (2006).

\bibitem{Navratil:10}
Petr Navr\'atil, Robert Roth, Sofia Quaglioni,
\newblock {\em Phys. Rev. C}
\newblock {\bf  82}, 034609 (2010).

\bibitem{Baroni:13prl}
Simone Baroni, Petr Navr\'atil, Sofia Quaglioni,
\newblock {\em Phys. Rev. Lett.}
\newblock {\bf  110}, 022505 (2013).

\bibitem{Raimondi:16}
F.~Raimondi, G.~Hupin, P.~Navr\'atil, S.~Quaglioni,
\newblock {\em Phys. Rev. C}
\newblock {\bf  93}, 054606 (2016).

\bibitem{Dickhoff:04}
W.H. Dickhoff, C.~Barbieri,
\newblock {\em Progress in Particle and Nuclear Physics}
\newblock {\bf  52}, 377  (2004).

\bibitem{Soma:11}
V.~Som\`a, T.~Duguet, C.~Barbieri,
\newblock {\em Phys. Rev. C}
\newblock {\bf  84}, 064317 (2011).

\bibitem{Barbieri:05}
C.~Barbieri, B.~K. Jennings,
\newblock {\em ibid.}
\newblock {\bf  72}, 014613 (2005).

\bibitem{Waldecker:11}
S.J. Waldecker, C.~Barbieri, W.~H. Dickhoff,
\newblock {\em ibid.}
\newblock {\bf  84}, 034616 (2011).

\bibitem{Carbone:13}
A.~Carbone, A.~Cipollone, C.~Barbieri, A.~Rios, A.~Polls,
\newblock {\em ibid.}
\newblock {\bf  88}, 054326 (2013).

\bibitem{Cipollone:13}
A.~Cipollone, C.~Barbieri, P.~Navr{\'a}til,
\newblock {\em Phys. Rev. Lett.}
\newblock {\bf  111}, 062501 (2013).

\bibitem{Cipollone:15}
A.~Cipollone, C.~Barbieri, P.~Navr\'atil,
\newblock {\em Phys. Rev. C}
\newblock {\bf  92}, 014306 (2015).

\bibitem{Ekstrom:15}
A.~Ekstr{\"o}m, {\em et al.},
\newblock {\em ibid.}
\newblock {\bf  91}, 051301 (2015).

\bibitem{Lapoux:16}
V.~Lapoux, V.~Som\`a, C.~Barbieri, H.~Hergert, J.~D. Holt, S.~R. Stroberg,
\newblock {\em Phys. Rev. Lett.}
\newblock {\bf  117}, 052501 (2016).

\bibitem{Garcia:16}
R.~F. Garcia~Ruiz, {\em et al.},
\newblock {\em Nat Phys}
\newblock {\bf  12}, 594 (2016).

\bibitem{Lister:66}
D.~Lister, A.~Sayres,
\newblock {\em Phys. Rev.}
\newblock {\bf  143}, 745 (1966).

\bibitem{Honore:86}
G.~M. Honor\'e, {\em et al.},
\newblock {\em Phys. Rev. C}
\newblock {\bf  33}, 1129 (1986).

\bibitem{Hagen:12}
G.~Hagen, N.~Michel,
\newblock {\em ibid.}
\newblock {\bf  86}, 021602 (2012).

\end{thebibliography}

\end{document}